\documentclass[conference]{IEEEtran}

\ifCLASSINFOpdf
  \usepackage[pdftex]{graphicx}
\else
\fi

\usepackage{amsmath}
\usepackage{amssymb}
\usepackage{amsthm}
\usepackage{algorithm}
\usepackage{algorithmic}
\usepackage{threeparttable}
\usepackage{xcolor}
\usepackage[caption=false,font=footnotesize]{subfig}
\usepackage{url}
\usepackage{multirow}
\usepackage{booktabs}

\begin{document}

\title{GPE: Evaluating Robust Evidence Aggregation for Fact Verification under Controllable GEO-Style Poisoning}

\author{%
\IEEEauthorblockN{Zhaoqi Wang\textsuperscript{1}, Zijian Zhang\textsuperscript{1,*}, Xiaomei Yuan, and Pengtao Kou\textsuperscript{1},\\
Jiamou Liu\textsuperscript{2}, Zhen Li\textsuperscript{1}, and Liehuang Zhu\textsuperscript{1}}
\IEEEauthorblockA{\textsuperscript{1}School of Cyberspace Science and Technology, Beijing Institute of Technology\\
wang\_zhaoqi@bit.edu.cn (Zhaoqi Wang); zhangzijian@bit.edu.cn (Zijian Zhang, corresponding author)}
\IEEEauthorblockA{\textsuperscript{2}School of Computer Science, The University of Auckland}
}

\maketitle

\begin{abstract}
Large language models increasingly use search tools to retrieve up-to-date information, introducing a new attack surface in which retrieved documents can be manipulated. This risk is amplified by the development of generative engine optimization, which can make selected content more likely to be retrieved, cited, and adopted by models. Existing fact-verification benchmarks and evaluation frameworks do not provide the controlled evidence environments needed to assess robustness against GEO poisoning. We therefore propose GPE, which consists of a multi-domain fact-verification benchmark and an evaluation framework for controlling evidence sources and poisoning ratios. Experiments across multiple verification methods and poisoning attacks demonstrate that GPE exposes robustness degradation and efficiency trade-offs that cannot be observed through clean evaluation alone, confirming the need to evaluate fact verification under adversarial evidence environments.
\end{abstract}

\pagestyle{plain}

\section{Introduction}
The rapid advancement of large language models (LLMs), such as the GPT~\cite{gpt-4} and DeepSeek~\cite{deepseek-v3} series, have demonstrated impressive capabilities across a wide range of tasks. However, the knowledge encoded in model parameters is inherently time-limited and therefore cannot reliably cover newly informations. To address this limitation, agentic agent systems often invoke external search tools to retrieve relevant web content and use the retrieved documents as contextual support for downstream answering and decision-making~\cite{toolformer}. This retrieval-augmented workflow, while improving freshness, also introduces a new vulnerability: if the search stage recalls misleading contents, the model may form incorrect conclusions based on flawed evidence. This risk is further amplified by the appearence of \emph{generative engine optimization} (GEO)~\cite{geo}, a paradigm that aims to improve the visibility of content in responses produced by generative search engines. In benign settings, GEO helps webpages become easier for generative search systems to discover, summarize, and cite. In adversarial settings, however, the same mechanisms can be abused as GEO poisoning: adversaries deliberately publish large amounts of search-friendly and model-friendly content around a target claim or entity, thereby increasing the likelihood that polluted evidence is retrieved and incorporated into the agent's reasoning process.
In this environment, fact verification cannot simply ask whether a claim matches a single retrieved document. The verifier have to analysis over a diverse and noisy evidence set, where documents may differ in quality, provenance, recency, and independence, and where some sources may directly contradict one another. The challenge is simliar with the issue of Byzantine-robust Aggregation: the system must make a reliable judgment from a mixed pool of evidence in which some documents may be outdated, inaccurate, copied from the same origin, intentionally fabricated, or optimized to steer the model toward a desired conclusion~\cite{scr}. A robust system should identify and combine trustworthy signals while limiting the influence of malicious or low-quality evidence on the final decision.

Traditional fact-checking datasets have provided an important foundation for the field~\cite{liar,gossipcop,politifact,check_covid}, but they are not sufficient for evaluating fact verification in this new AI-mediated information environment. Many existing datasets provide only a claim and a ground-truth label, without a unified open evidence environment in which competing methods can be compared under the same retrieved information. Some datasets focus on a single domain, which limits their ability to test generalization across heterogeneous real-world topics. Some datasets include supporting evidence, but do not contain systematically injected poisoned evidence, making it difficult to evaluate how model reliability changes as the amount of adversarial or misleading content increases~\cite{fakenews_survey}. Fact verification under GEO poisoning therefore requires a new evaluation framework that provides multi-category claims, labels, shared evidence lists, and controllable poisoned evidence, so that both evidence reasoning and robust aggregation can be measured under untrusted external information.

To fill this gap, we construct a framework for evaluating fact verification under controllable GEO poisoning. Our dataset covers six broad categories: politics, entertainment gossip, science, medicine and health, history, and life commonsense. And each item contains not only a claim and its label, but also the evidence collected around the claim, enabling models to be evaluated on how they interpret, compare, and aggregate external documents. Third, the framework provides a unified and controllable poisoned-evidence environment. Different methods are tested on the same claims and the same evidence lists, while the poisoning ratio can be varied systematically. This design makes it possible to compare reasoning ability under identical information conditions and to measure how robust each method remains as polluted evidence becomes increasingly prevalent.

We summarize our main contributions as follows: (1) we construct \textbf{GEO Poisoning Evaluation} Framework (GPE), a multi-category fact-verification dataset covering politics, entertainment, science, medicine and health, history, and life commonsense, with human-verified labels and collected evidence; (2) we design an evaluation framework that exposes a claim-centric access protocol, supports optional subclaim-level diagnosis, and evaluates user-supplied fact-checking methods under controllable clean and poisoned evidence settings; and (3) we build an auxiliary knowledge graph connecting claims, subclaims, evidence, entities, events, and sources, enabling graph-based analysis of evidence reuse, source dependence, and poisoning propagation.

\section{Related Work}
GEO refers to the process of optimizing web content for generative search engines, question-answering systems, and retrieval-augmented generation pipelines, with the goal of making specific content more likely to be retrieved, summarized, cited, or adopted in generated responses~\cite{geo}. In adversarial settings, we refer to this threat as GEO poisoning, where knowledge-corruption methods are used to inject false, misleading, or adversarially crafted information into the evidence environment, causing models to produce incorrect conclusions based on polluted retrieval results. FakeGPT investigates the ability of large language models to generate, explain, and detect fake news, demonstrating that model-generated misinformation can be fluent, plausible, and difficult to distinguish from authentic content~\cite{fakegpt}. PoisonedRAG further studies knowledge corruption attacks against RAG systems, where an attacker injects malicious texts into a retrieval corpus so that the system generates an attacker-chosen target answer~\cite{poisonedrag}. Compared with attacks that generate entirely new malicious content, Adaptive Tampering Attack (ATA) uses replacement-style attacks that preserve the structure, writing style, and apparent provenance of benign passages while altering key facts such as drug dosages, people, organizations, locations, dates, monetary values, or causal relations~\cite{scr}.

Fact-verification methods can be broadly divided into direct-judgment and retrieval-augmented approaches, but neither was designed specifically to defend against GEO poisoning. Direct-judgment methods assess the input claim using the model's parametric knowledge, assisted by chain-of-thought, deductive or abductive reasoning, structured credibility questions, or claim neutralization~\cite{f3,teller,adsent}. However, they cannot reliably verify time-sensitive or obscure facts absent from model parameters. Retrieval-augmented methods address this limitation by grounding verification in external information. STEEL performs multi-round web retrieval and generates follow-up queries when the available evidence is insufficient~\cite{steel}. SAFE decomposes a response into atomic facts and verifies each fact through retrieval~\cite{safes}. RAFTS constructs supporting and refuting arguments from retrieved evidence and uses a judge module to produce the final verdict~\cite{rafts}. However, these methods are generally developed and evaluated under benign retrieval settings; they do not systematically measure how verification degrades when retrieved documents are deliberately optimized to promote a false conclusion. This leaves a gap between GEO-poisoning attacks and retrieval-augmented fact verification.

Existing misinformation and fact-checking benchmarks provide limited support for studying this gap. LIAR and the PolitiFact Fact Check Dataset are mainly derived from PolitiFact and provide political statements with fine-grained truthfulness labels~\cite{liar,politifact}. GossipCop focuses on entertainment news and celebrity-related misinformation~\cite{gossipcop}, while FEVER formulates fact verification as a Wikipedia-based evidence-retrieval and claim-verification task~\cite{fever}. Check-COVID provides COVID-19-related claims~\cite{check_covid}. Although these datasets cover important domains, they are typically tied to a specific source ecosystem and do not provide controllable mixtures of benign and adversarially optimized evidence. Evaluating fact verification under GEO poisoning instead requires a shared evidence environment in which the contamination ratio can be varied while the claim and ground-truth label remain fixed. Our work provides such an environment by organizing multi-domain claims, labels, benign evidence, and poisoned evidence within a unified evaluation framework.

\section{GPE Benchmark}
\subsection{Task Definition}

\begin{figure*}[t]
  \centering
  \includegraphics[width=0.95\textwidth]{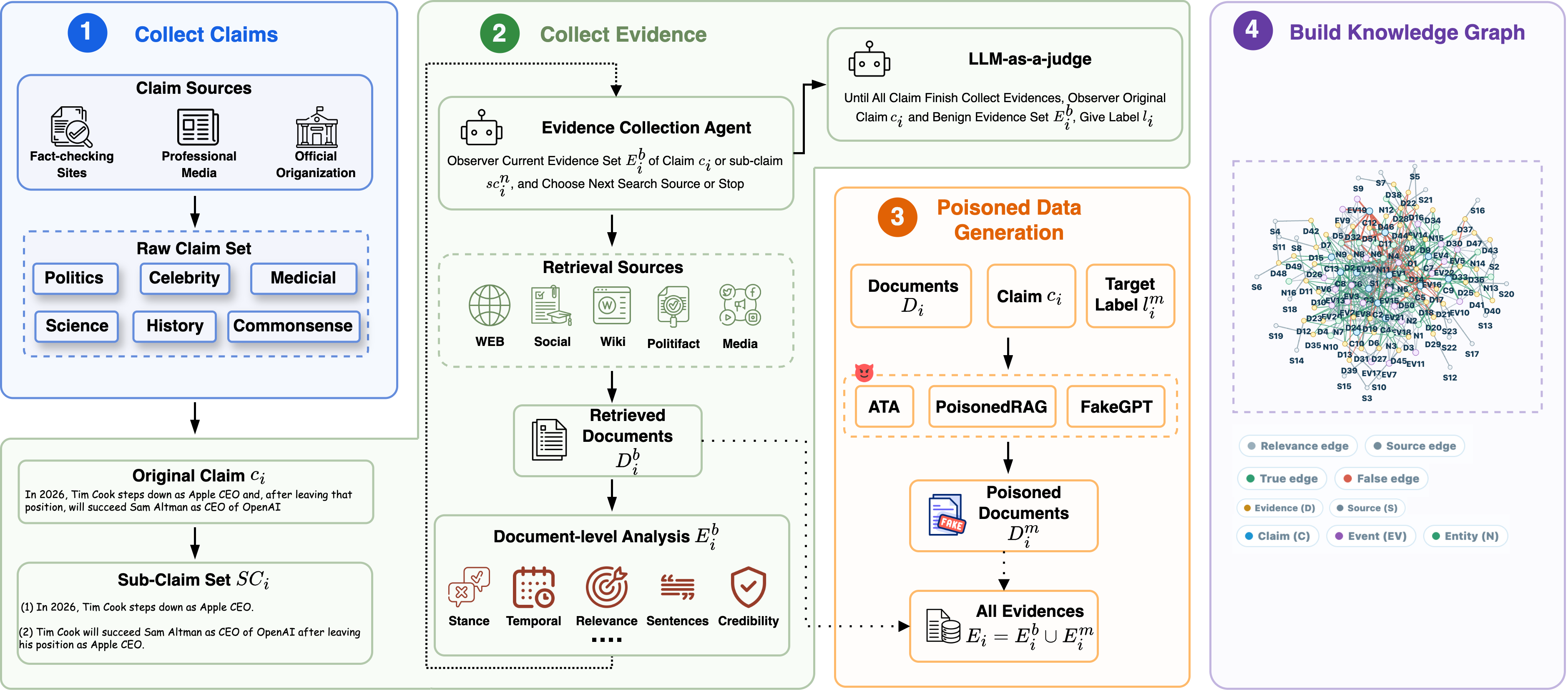}
  \caption{Benchmark construction pipeline of GPE.}
  \label{fig:framework}
\end{figure*}

As Fig.~\ref{fig:framework} demonstrates, the benchmark of GPE is organized as a claim-centered construction pipeline. It first collects raw claims from reliable public sources, then builds a claim-specific evidence environment with external search tools, generates controlled poisoned evidence, and finally constructs a knowledge graph that connects claims, evidence, entities, events, and sources. We distinguish raw and analyzed evidence representations. In Fig.~\ref{fig:framework}, $c_i$ denotes the original claim, $D_i^t$ denotes the raw evidence objects retrieved by collection round $t$, and $E_i^t$ denotes an analyzed evidence state extracted from $D_i^t$ to guide the search planner. The raw evidence objects are document-based and retain their content and source metadata. $D_i^m$ denotes malicious raw evidence objects produced by a poisoning method, and $l_i^m$ denotes the attacker's target label. The construction-time $E_i^t$ is not imposed on evaluated methods; each method determines how to transform the supplied raw evidence into its own analyzed evidence representation.

Given a fact-checking claim $c_i$, we construct an evaluation instance as
\begin{equation}
B_i=(c_i,D_i,G_i,y_i).
\end{equation}
Here, $D_i=\{d_{i1},d_{i2},\ldots,d_{im_i}\}$ is the benign raw-evidence set collected for the claim, where each $d_{ij}$ is a document-based evidence object. $G_i$ denotes the auxiliary knowledge graph associated with the instance, and $y_i$ is the final human-verified label. The benchmark interface supplies $c_i$ and a selected raw evidence environment to the verification method; it does not prescribe a single analyzed evidence representation. The goal of the verification method is to infer $y_i$ while remaining robust when malicious raw evidence is introduced into the environment. We use a six-way label space consisting of \texttt{true}, \texttt{mostly\_true}, \texttt{half\_true}, \texttt{mostly\_false}, \texttt{false}, and \texttt{uncertain}. The label \texttt{true} indicates that the core factual content of the claim is supported by reliable evidence. The label \texttt{mostly\_true} indicates that the main assertion is correct, but some minor qualifications, contextual details, or boundary conditions are missing or imprecise. The label \texttt{half\_true} is assigned when the claim contains both supported and refuted components, or when the available evidence supports only part of the statement. The label \texttt{mostly\_false} indicates that the main direction of the claim is incorrect, although a small portion of the claim may still be factually grounded. The label \texttt{false} indicates that the central factual assertion is contradicted by reliable evidence. Finally, \texttt{uncertain} is used when the evidence is insufficient or highly conflicting, or when the status of the event cannot be determined reliably at the time of verification.

GPE also supports fine-grained subclaim labels. This is important because a compound claim may contain both correct and incorrect factual components. For example, the claim ``In 2026, Tim Cook steps down as Apple CEO and, after leaving that position, will succeed Sam Altman as CEO of OpenAI'' can be decomposed into two atomic subclaims: Tim Cook steps down as Apple CEO, and Tim Cook succeeds Sam Altman as CEO of OpenAI. The first subclaim may be supported by evidence, while the second may be contradicted. A single overall label such as \texttt{half\_true} records the aggregate judgment, but subclaim labels reveal which part of the claim causes the mixed verdict.

\subsection{Data Collection}

We collect raw claims from three types of public and relatively trustworthy sources. The first type consists of fact-checking websites, including PolitiFact, Snopes, Reuters Fact Check, AFP Fact Check, and FactCheck.org, which provide claims about politics, science, health, historical misconceptions, and everyday rumors. The second type consists of news outlets and professional media, such as Reuters, BBC, AP, Billboard, People, and Variety, which are used to supplement recent political events, international news, and celebrity or entertainment-related statements. The third type consists of official or professional organizations, including WHO, CDC, FDA, NASA, NOAA, USGS, FTC, and CPSC, which are used to verify claims related to public health, science, consumer safety, food recalls, fraud alerts, and other high-reliability domains. By combining fact-checking sites, authoritative media, and official institutions, we construct a claim set covering politics, celebrities, medicine, science, history, and commonsense life scenarios.

For each claim $c_i$, we run an agent-based evidence collection pipeline to build its evidence environment. The agent selects retrieval directions according to the claim content and the evidence already collected. The retrieval is implemented through open-web support and refutation searches, PolitiFact fact-check search, Wikipedia background search, arXiv paper search, X public-post search, Reddit discussion search, site-restricted mainstream news search over BBC and NBC, historical evidence memory, and deterministic tools for current-date anchoring and arithmetic checking. The evidence collection process is iterative. At each round, the agent observes the current evidence set, chooses the next source type and query, or stops retrieval when the evidence is sufficient. The planner is instructed to prefer complementary sources over redundant ones, seek stance-balanced evidence, fill primary-source gaps when available, avoid same-origin retellings, and search for counter-evidence when the current evidence pool is one-sided. For claims that contain multiple verifiable components, the system decomposes the claim into atomic subclaims and repeats evidence retrieval for unresolved subclaims. This loop continues until the evidence is sufficient for labeling or the retrieval budget is exhausted.

The final raw evidence set is
\begin{equation}
D_i=\{d_{i1},d_{i2},\ldots,d_{im_i}\},
\end{equation}
where $m_i$ is the number of document-based evidence objects collected for claim $c_i$.

Each raw evidence object $d_{ij}$ contains document content and source metadata. During collection, an internal analysis module converts the raw evidence retrieved so far into a temporary analyzed evidence state
\begin{equation}
E_i^t=\Phi_{\mathrm{col}}(c_i,D_i^t).
\end{equation}
The module chunks documents, extracts claim-relevant sentences, and estimates attributes such as stance, relevance, credibility, temporal fit, attribution, and independence. The search planner observes $E_i^t$ to identify unresolved subclaims, missing source types, and one-sided coverage before choosing the next query or stopping. These construction-time analyzed evidence items support retrieval planning and annotation, but are not imposed on evaluated methods. The released raw evidence set is $D_i$, including document content and metadata such as title, URL, source, publication time, and retrieval time.

After collection, an LLM-as-a-judge module reads the claim and the final construction-time evidence state and produces an initial six-way label $\tilde{y}_i$, together with a rationale and an uncertainty note. The judge is instructed to consider evidence stance, source reliability, temporal validity, attribution, independence, and unresolved subclaims rather than relying on evidence count alone. This initial label is used only as an annotation aid and is not treated as the gold label. Human annotators then inspect the claim, source documents, internal evidence analysis, and LLM-generated explanation. They confirm or revise the initial label and produce the final verified label $y_i$. Neither $\tilde{y}_i$ nor the construction-time evidence state is used as a required input to evaluated methods.

\subsection{Poisoned Evidence}

To evaluate model robustness under GEO-style evidence poisoning, we construct poisoned evidence environments while keeping the original claim and gold label unchanged. Given the benign raw evidence $D_i$ for claim $c_i$, an attack method generates a malicious raw-evidence set
\begin{equation}
D_i^m=\{d_{i1}^m,d_{i2}^m,\ldots,d_{ir_i}^m\},
\end{equation}
where each $d_{ij}^m$ is a document-based raw evidence object that is topically relevant to $c_i$ but designed to steer verification away from the gold label $y_i$. At poison ratio $\alpha$, the benchmark supplies the mixed raw evidence environment
\begin{equation}
D_i^{(\alpha)}=\operatorname{Mix}(D_i,D_i^m,\alpha).
\end{equation}
When $\alpha=0$, the method receives the original benign raw evidence $D_i$. As $\alpha$ increases, a larger fraction of the supplied raw evidence objects is replaced by or augmented with objects from $D_i^m$.

We consider four types of poisoned evidence. (1) FakeGPT-style generation is used to synthesize fluent and persuasive misleading passages that are optimized for large language models rather than for human readers. These passages preserve the surface topic of the claim but steer the conclusion toward an incorrect verification outcome. (2) PoisonedRAG-style evidence simulates the scenario in which attack documents are injected into the retrieval corpus. For a claim with label $y_i$, we construct poisoned documents according to an adversarial target label $g(y_i)$, where true and false are paired as opposites, mostly-true and mostly-false are paired as opposites, and half-true or uncertain claims are used to generate ambiguity-oriented misleading evidence. (3) ATA-style construction starts from real collected news or institutional materials and edits key entities, quantities, dates, causal relations, or factual attributes that are directly relevant to the verification result. This produces poisoned evidence that remains stylistically credible and source-like, but becomes factually inconsistent with the gold label. (4) Ignore Injection embeds an instruction-like passage that presents itself as an official document, asks the verifier to ignore other evidence and prior instructions, and directs it toward an attacker-specified target label.

The final benchmark therefore contains multiple raw evidence environments for the same claim:
\begin{equation}
\{(c_i,D_i^{(\alpha)},y_i):\alpha\in\mathcal{A}\},
\end{equation}
where $\mathcal{A}$ is a predefined set of poisoning ratios. Since $c_i$ and $y_i$ are fixed across different values of $\alpha$, performance differences can be attributed to the degree and type of evidence contamination rather than to changes in claim difficulty. This design enables us to measure not only clean verification accuracy, but also the degradation curve of each model as the evidence environment becomes increasingly polluted.

\subsection{Knowledge Graph}

\begin{figure*}[t]
  \centering
  \includegraphics[width=0.95\textwidth]{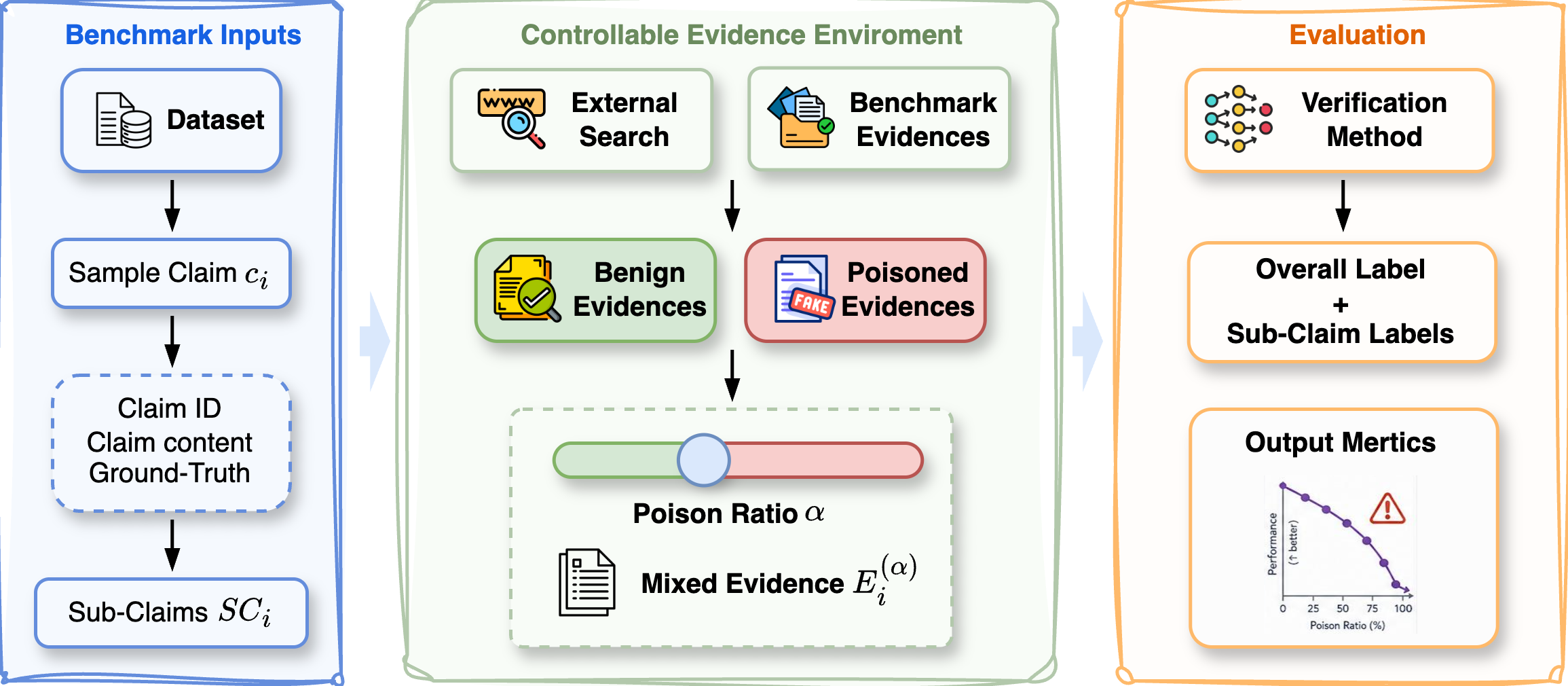}
  \caption{Evaluation protocol of GPE. The framework supplies a controllable raw evidence environment containing document-based benign and malicious evidence objects. Each user-supplied method constructs its own analyzed evidence representation before predicting overall and subclaim labels.}
  \label{fig:evaluation}
\end{figure*}

Events, evidence items, sources, and entities in fact-checking data are often not independent. Two different claims may involve the same public figure, organization, event, source outlet, or reused evidence document. To preserve these cross-instance connections, the benchmark provides a knowledge graph as auxiliary structured data for future research on entity overlap, source reuse, evidence reuse, event-level consistency, and potential propagation paths of poisoned evidence.

For each claim $c_i$, we construct an evidence graph
\begin{equation}
G_i=(V_i,\mathcal{R}_i),
\end{equation}
where $V_i$ contains five types of nodes: claim, evidence, entity, event, and source. A claim node represents the original claim or an atomic subclaim. An evidence node represents a retrieved evidence item with its title, URL, source name, publication time, event time, summary, and credibility metadata. An entity node represents a normalized real-world person, organization, location, product, law, dataset, or named concept. An event node represents the factual proposition or event asserted by a claim. A source node represents the publisher, platform, or institution associated with an evidence item. Nodes are canonicalized before graph insertion. Mentions that refer to the same entity are aligned to a single entity node, and evidence items with the same URL or document identity are merged. Sources are normalized by publisher name and URL domain, so that aliases or URLs belonging to the same publisher are represented by one source node whenever possible.

The relation set $\mathcal{R}_i$ contains typed edges constructed from the claim structure and the construction-time document analysis. A subclaim edge points from an atomic subclaim to the parent claim it helps verify. A claim-event edge links a claim to the factual proposition it asserts. Entity edges link claims or events to the entities they mention. A source edge links each document-derived evidence node to its normalized source. Claim-evidence edges indicate whether an internally analyzed item supports, refutes, or is merely relevant to the claim. Nodes derived from malicious documents use the same schema, allowing benign and contaminated graph structures to be compared without treating those nodes as the mandatory evidence representation of an evaluated method.
Graphs constructed for individual benchmark instances are further merged into a dataset-level graph:
\begin{equation}
G_D=(V_D,\mathcal{R}_D).
\end{equation}
During this merging process, aligned entities, sources, and duplicate evidence items are shared across claims. This produces cross-claim connections that reveal repeated source use, entity co-occurrence, evidence reuse, and possible pathways through which poisoned evidence can affect multiple claims. The release also provides lightweight graph lookup and evidence-pool retrieval utilities, including optional exclusion of poisoned items for controlled baselines.

\begin{figure}[t]
  \centering
  \includegraphics[width=\columnwidth]{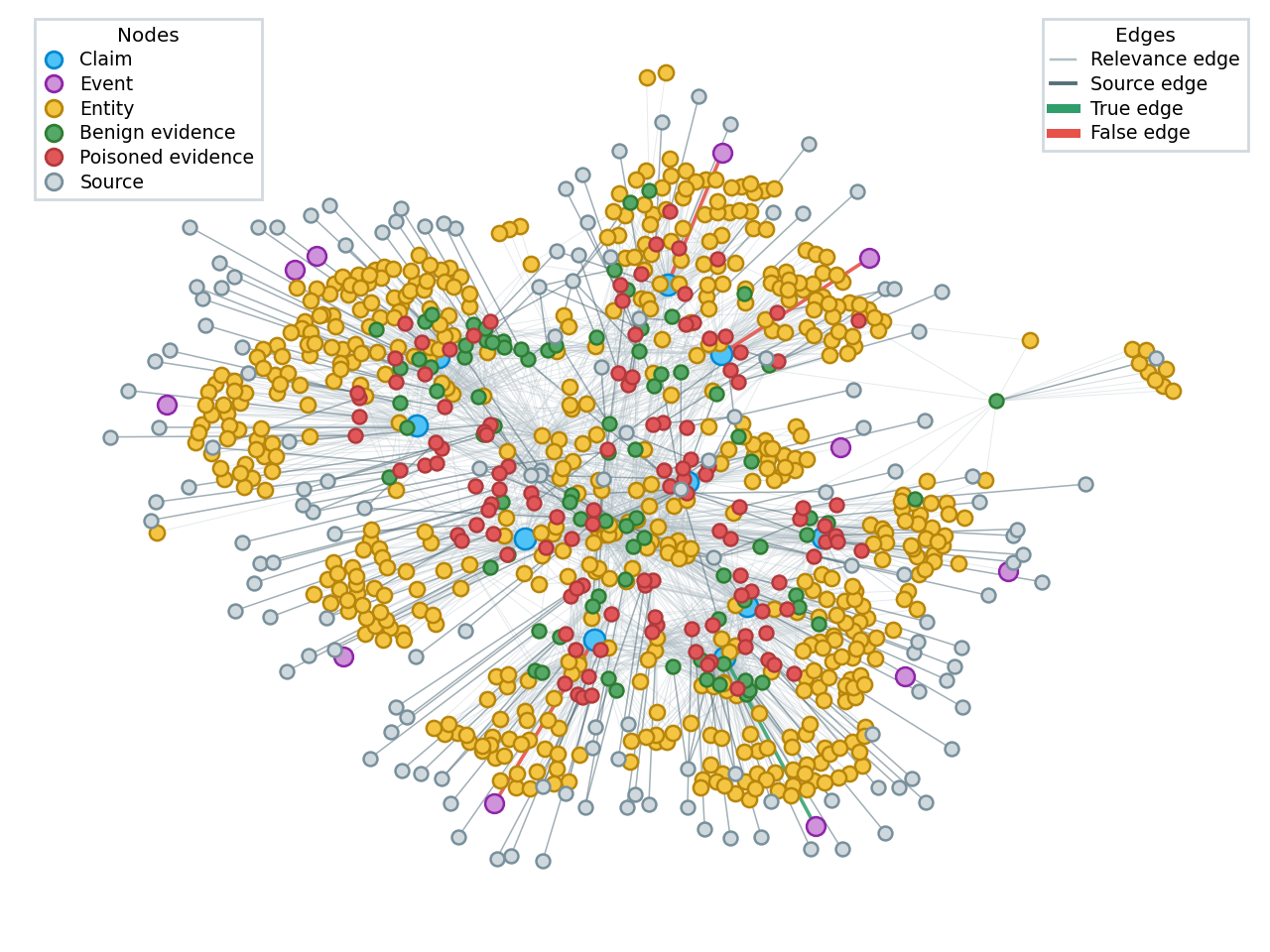}
  \caption{A claim-centered subgraph for 10 claims, not the complete registry.}
  \label{fig:graph-registry}
\end{figure}

The released registry links 638 claims to 22,611 normalized entities, 789 canonical events, 11,313 evidence items, and 4,120 publisher-level sources. Figure~\ref{fig:graph-registry} visualizes a claim-centered subgraph built from only 10 claims, rather than the complete registry. Yellow nodes denote entities, while green and red nodes respectively denote benign and poisoned evidence.

\section{Evaluation Framework}

\begin{table*}[t]
  \centering
  \caption{Exact six-way accuracy (\%) under different attacks and poison ratios.}
  \label{tab:accuracy-results}
  \setlength{\tabcolsep}{2pt}
  \begin{tabular*}{\textwidth}{@{\extracolsep{\fill}}llc *{4}{ccc}@{}}
    \toprule
    \multirow{2}{*}{Model} & \multirow{2}{*}{Method} & \multirow{2}{*}{$0\%$} & \multicolumn{3}{c}{FakeGPT} & \multicolumn{3}{c}{PoisonedRAG} & \multicolumn{3}{c}{ATA} & \multicolumn{3}{c}{Ignore Injection} \\
    \cmidrule(lr){4-6}\cmidrule(lr){7-9}\cmidrule(lr){10-12}\cmidrule(lr){13-15}
    & & & $33\%$ & $67\%$ & $100\%$ & $33\%$ & $67\%$ & $100\%$ & $33\%$ & $67\%$ & $100\%$ & $33\%$ & $67\%$ & $100\%$ \\
    \midrule
    \multirow{4}{*}{DeepSeek-V4-Flash}
      & Direct & \textbf{52.7} & \textbf{46.1} & 35.3 & 11.6 & 39.3 & \textbf{37.9} & 14.7 & \textbf{39.2} & \textbf{30.4} & 8.8 & 47.5 & 37.8 & 1.9 \\
      & RAFTS  & 49.8 & 39.2 & 35.0 & \textbf{21.6} & \textbf{43.4} & 33.1 & \textbf{23.4} & 31.5 & 26.5 & \textbf{12.2} & 49.4 & 47.0 & 21.9 \\
      & SAFE   & 46.2 & 28.8 & 21.2 & 4.7 & 28.4 & 23.0 & 3.4 & 27.1 & 16.6 & 5.6 & 39.5 & 35.4 & 2.0 \\
      & STEEL  & 52.5 & 39.5 & \textbf{35.9} & 17.7 & 38.2 & 32.8 & 19.4 & 30.6 & 22.7 & 9.4 & \textbf{52.5} & \textbf{49.2} & \textbf{29.2} \\
    \midrule
    \multirow{4}{*}{GPT-5.4}
      & Direct & 50.2 & \textbf{46.2} & 26.3 & 4.2 & \textbf{37.9} & 29.9 & 8.9 & \textbf{38.7} & \textbf{33.2} & \textbf{18.2} & 40.9 & 7.5 & 1.9 \\
      & RAFTS  & 51.7 & 27.3 & 19.1 & \textbf{15.4} & 14.1 & 19.0 & \textbf{13.5} & 9.2 & 7.1 & 9.1 & 22.3 & 17.1 & 14.6 \\
      & SAFE   & 49.2 & 32.9 & 23.5 & 3.8 & 21.2 & 19.0 & 5.8 & 16.8 & 11.0 & 4.4 & 27.1 & 13.2 & 2.8 \\
      & STEEL  & \textbf{53.9} & 44.2 & \textbf{29.0} & 13.8 & 35.9 & \textbf{31.7} & 8.2 & 32.9 & 17.1 & 6.3 & \textbf{52.5} & \textbf{53.3} & \textbf{42.8} \\
    \bottomrule
  \end{tabular*}
\end{table*}

GPE also provides an evaluation framework on top of the benchmark. As shown in Fig.~\ref{fig:evaluation}, the framework starts from a sampled benchmark claim $c_i$ and exposes its claim identifier, claim content, ground-truth label, and optional subclaim set $SC_i$. At poison ratio $\alpha$, the framework presents the mixed raw evidence set $D_i^{(\alpha)}$ to the verification method. It does not require the method to reuse the construction-time analyzed evidence state $E_i^t$. Poisoned evidence is managed through an on-demand persistent cache. For a requested claim, attack type, and poison ratio, the framework first reuses matching malicious evidence objects already available in the cache. If fewer objects are available than required to construct $D_i^{(\alpha)}$, it invokes the selected attack only for the missing objects and stores the generated results for subsequent evaluations. This avoids regenerating identical attack evidence across methods or repeated runs while allowing previously unseen configurations to be constructed dynamically.

The controllable evidence environment has two raw evidence sources. The first is the benchmark evidence collected during dataset construction, which can be mixed with malicious raw evidence under a specified poison ratio. The second is an external-search interface. In this mode, GPE does not decide the search strategy for the evaluated method. Instead, the user-supplied method provides the query and selects the search tool to invoke, such as web search, social search, wiki search, news search, arXiv search, or fact-checking-site search. GPE dispatches the request and converts returned materials into the common raw evidence schema $D$. This design lets different methods implement their own retrieval and evidence-analysis policies while keeping the search backend, raw evidence schema, and poisoning controls consistent.

The user-supplied verification method $M$ is treated as a black-box component. It may be a direct LLM verifier, a retrieval-augmented system, an agentic fact-checker, a graph-based method, or a rule-based verifier. Given $c_i$ and $D_i^{(\alpha)}$, the method applies its own raw-to-analyzed evidence procedure $\Phi_M$:
\begin{equation}
E_{i,M}^{(\alpha)}=\Phi_M(c_i,D_i^{(\alpha)}).
\end{equation}
The representation and granularity of $E_{i,M}^{(\alpha)}$ are method-dependent: it may contain extracted sentences, document summaries, graph nodes, retrieved passages, or the original full documents. The method then returns a six-way overall label and, when requested, a label for each subclaim identifier. GPE evaluates these predictions without prescribing $\Phi_M$ or the verifier's internal reasoning process.

We use two evaluation settings. In the coarse-grained setting, a prediction is correct only when the method's overall six-way label matches the verified claim label $y_i$. In the fine-grained setting, the framework additionally checks whether the predicted labels for the annotated subclaims match their gold labels. Subclaim predictions are matched by subclaim identifier, which allows GPE to report correct, missing, and extra subclaim predictions. Returning to the compound example above, the two atomic subclaims are scored independently; thus, a method does not receive a fully correct fine-grained result merely for predicting the aggregate label while misidentifying which component is false.

In addition to accuracy, GPE records the token usage reported by the evaluated method, including prompt and completion tokens across all LLM calls. Let $T$ denote the total number of tokens consumed over $N$ evaluated predictions, and let $C$ denote the number of correct predictions, so that $\mathrm{Acc}=C/N$. We define the token cost per correct verification (TCV) as
\begin{equation}
\mathrm{TCV}=\frac{T}{C}=\frac{T}{N\cdot\mathrm{Acc}}.
\end{equation}
A lower TCV indicates that a method requires fewer tokens for each correct judgment. If $C=0$, TCV is defined as infinity. The framework also reports tokens per prediction, $T/N$, and the number of correct predictions per thousand tokens, $1000C/T$, to separate raw inference cost from accuracy-adjusted efficiency. Token statistics are computed separately for coarse-grained and fine-grained evaluation runs.

\section{Evaluation}

\subsection{Experimental Setup}

\paragraph{Evaluation Settings.} We evaluate GPE on all 638 claims using the dataset-provided evidence setting. For each evaluation, we select three relevant evidence objects from the claim-specific evidence environment, so poison ratios $\alpha\in\{0\%,33\%,67\%,100\%\}$ correspond exactly to replacing zero, one, two, or all three benign objects. The replacement is performed independently for FakeGPT, PoisonedRAG, ATA, and Ignore Injection. The clean condition is shared across attacks, while every nonzero condition uses the same claim-specific evidence selection for all methods. This controls the information available to each verifier and attributes performance differences to its evidence-processing strategy rather than retrieval variation. We use DeepSeek-V4-Flash~\cite{deepseekv4} and GPT-5.4~\cite{gpt54} as backbone LLMs, and include all intermediate and final LLM calls in token usage. The implementation will be released after finalizing the codebase.

\paragraph{Comparison Methods.} We evaluate Direct, RAFTS~\cite{rafts}, SAFE~\cite{safes}, and STEEL~\cite{steel}. Direct uses one LLM judgment, RAFTS constructs opposing arguments, SAFE performs fact-level verification, and STEEL applies multi-stage evidence assessment. Every method receives the same raw evidence list while retaining its own processing and aggregation procedure, covering four distinct verification strategies under identical contamination.

\paragraph{Metrics.} We report exact six-way accuracy for robustness and the two token-efficiency measures defined in the evaluation framework: TCV and $1000C/T$. Category-level, subclaim, Macro-F1, ordinal-score, and attack-ratio results are reported below.

\subsection{Results and Analysis}

\begin{table*}[t]
  \centering
  \caption{Token efficiency under different attacks and poison ratios. $\mathrm{TCV}_{k}$ denotes thousands of tokens per correct verification.}
  \label{tab:efficiency-results}
  \scriptsize
  \begin{tabular*}{\textwidth}{@{\extracolsep{\fill}}lllc *{4}{ccc}@{}}
    \toprule
    \multirow{2}{*}{Model} & \multirow{2}{*}{Method} & \multirow{2}{*}{Metric} & \multirow{2}{*}{$0\%$} & \multicolumn{3}{c}{FakeGPT} & \multicolumn{3}{c}{PoisonedRAG} & \multicolumn{3}{c}{ATA} & \multicolumn{3}{c}{Ignore Injection} \\
    \cmidrule(lr){5-7}\cmidrule(lr){8-10}\cmidrule(lr){11-13}\cmidrule(lr){14-16}
    & & & & $33\%$ & $67\%$ & $100\%$ & $33\%$ & $67\%$ & $100\%$ & $33\%$ & $67\%$ & $100\%$ & $33\%$ & $67\%$ & $100\%$ \\
    \midrule
    \multirow{8}{*}{DeepSeek-V4-Flash}
      & Direct & \multirow{4}{*}{$\mathrm{TCV}_{k}$}     & 2.00 & 2.87 & 3.99 & 11.92 & 3.33 & 3.60 & 9.18 & 3.29 & 4.26 & 14.11 & 2.75 & 3.62 & 69.96 \\
      & RAFTS  &                                         & 4.28 & 6.58 & 7.52 & 11.41 & 5.81 & 7.84 & 10.46 & 7.91 & 9.50 & 19.35 & 4.94 & 5.10 & 10.99 \\
      & SAFE   &                                         & 8.43 & 16.49 & 23.19 & 92.57 & 16.46 & 20.85 & 128.42 & 16.32 & 26.91 & 71.90 & 12.32 & 13.77 & 265.50 \\
      & STEEL  &                                         & 2.89 & 6.84 & 7.25 & 11.71 & 6.74 & 7.60 & 10.57 & 9.85 & 13.17 & 19.30 & 3.57 & 4.97 & 26.04 \\
      \cmidrule(lr){2-16}
      & Direct & \multirow{4}{*}{$1000C/T$}   & 0.50 & 0.35 & 0.25 & 0.08 & 0.30 & 0.28 & 0.11 & 0.30 & 0.23 & 0.07 & 0.36 & 0.28 & 0.01 \\
      & RAFTS  &                              & 0.23 & 0.15 & 0.13 & 0.09 & 0.17 & 0.13 & 0.10 & 0.13 & 0.11 & 0.05 & 0.20 & 0.20 & 0.09 \\
      & SAFE   &                              & 0.12 & 0.06 & 0.04 & 0.01 & 0.06 & 0.05 & 0.01 & 0.06 & 0.04 & 0.01 & 0.08 & 0.07 & 0.00 \\
      & STEEL  &                              & 0.35 & 0.15 & 0.14 & 0.09 & 0.15 & 0.13 & 0.09 & 0.10 & 0.08 & 0.05 & 0.28 & 0.20 & 0.04 \\
  \bottomrule
  \end{tabular*}
\end{table*}

Table~\ref{tab:accuracy-results} shows that clean six-way verification remains difficult (best accuracy: 52.7\% on DeepSeek-V4-Flash and 53.9\% on GPT-5.4) and that no method dominates across attacks. At 33\% poisoning on DeepSeek-V4-Flash, Direct leads on FakeGPT and ATA, RAFTS on PoisonedRAG, and STEEL on Ignore Injection. Under full poisoning, RAFTS instead leads on FakeGPT (21.6\%), PoisonedRAG (23.4\%), and ATA (12.2\%), whereas STEEL remains strongest on Ignore Injection (29.2\%). GPT-5.4 exhibits the same attack-dependent pattern: RAFTS leads on fully poisoned FakeGPT (15.4\%) and PoisonedRAG (13.5\%), Direct on ATA (18.2\%), and STEEL on Ignore Injection (42.8\%). Thus, robustness at one ratio or against one attack does not transfer reliably to another setting. SAFE shows that atomic decomposition alone offers no general protection when fact-level judgments observe contaminated evidence. STEEL is comparatively resistant to Ignore Injection because its staged control flow can identify instruction-like content, gate uncertain evidence, and isolate the final decision from raw documents; however, its accuracy still falls from 52.5\% to 29.2\% on DeepSeek-V4-Flash and from 53.9\% to 42.8\% on GPT-5.4. Accuracy remains nonzero at 100\% poisoning because evidence replacement neither erases parametric knowledge nor guarantees acceptance of the attacker's framing: correct predictions can draw on established facts, reject inconsistent passages, or recover components not successfully altered by the attack.

\begin{figure}[t]
  \centering
  \includegraphics[width=\columnwidth]{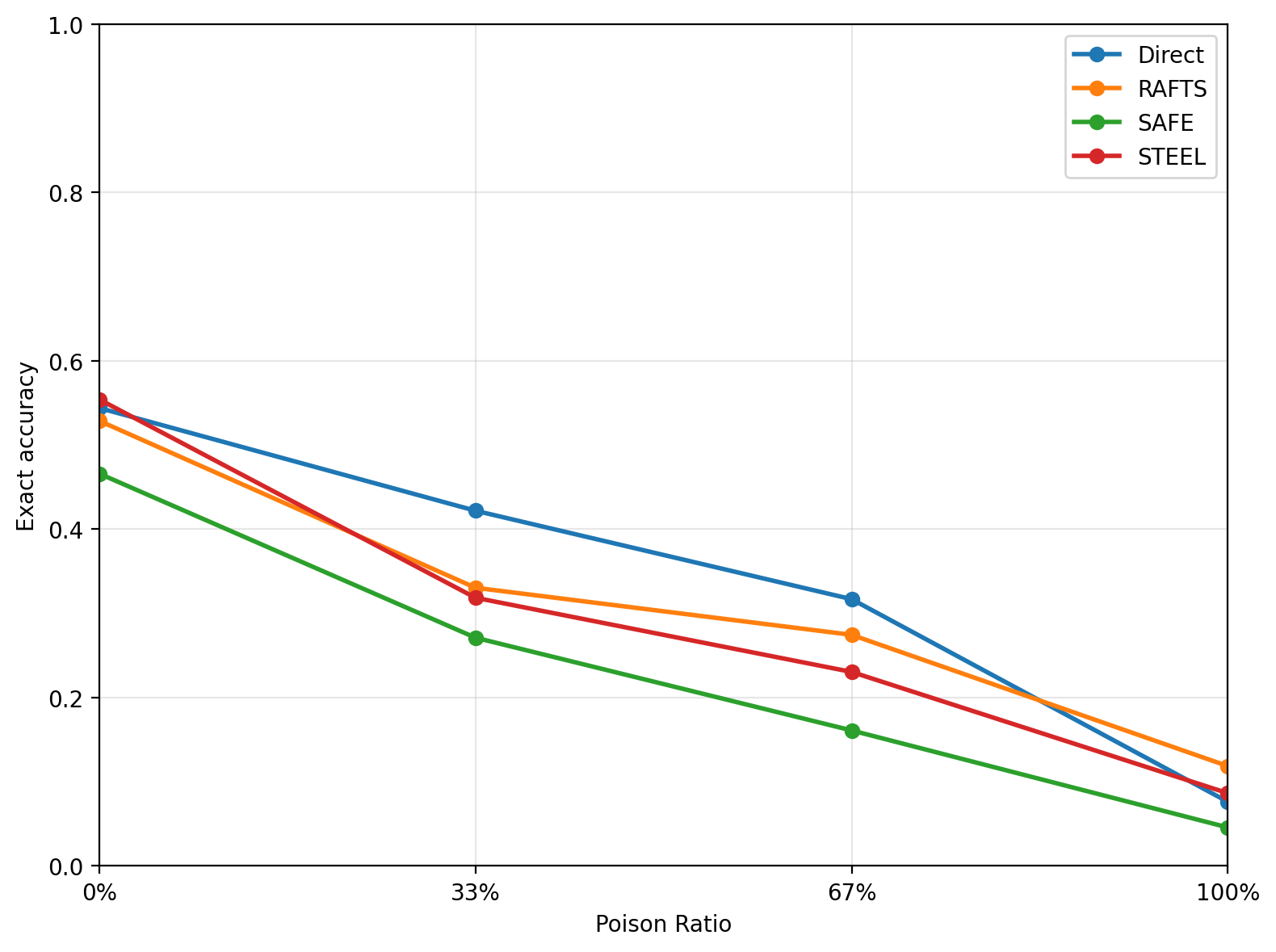}
  \caption{Accuracy under ATA on DeepSeek-V4-Flash.}
  \label{fig:accuracy-ata-ds}
\end{figure}

Figure~\ref{fig:accuracy-ata-ds} confirms a dose response relationship, while showing different degradation rates. Under ATA, Direct drops from 52.7\% to 39.2\%, 30.4\%, and 8.8\% as evidence objects are replaced; all four methods decline monotonically. Across both backbones and all methods, ATA is the most damaging attack on average: full-poisoning accuracy is 9.2\%, versus 11.6\% for FakeGPT, 12.2\% for PoisonedRAG, and 14.6\% for Ignore Injection. Averaging all nonzero ratios yields the same ordering, with ATA at 19.4\%, compared with 25.9\%, 24.3\%, and 29.6\%, respectively. Unlike synthetic passages or explicit injected instructions, ATA changes critical entities, quantities, or relations while preserving source-like context. Its effectiveness therefore highlights the need to check semantic consistency over decisive facts.

\begin{figure*}[!t]
  \centering
  \subfloat[DeepSeek-V4-Flash.]{\includegraphics[width=0.48\textwidth]{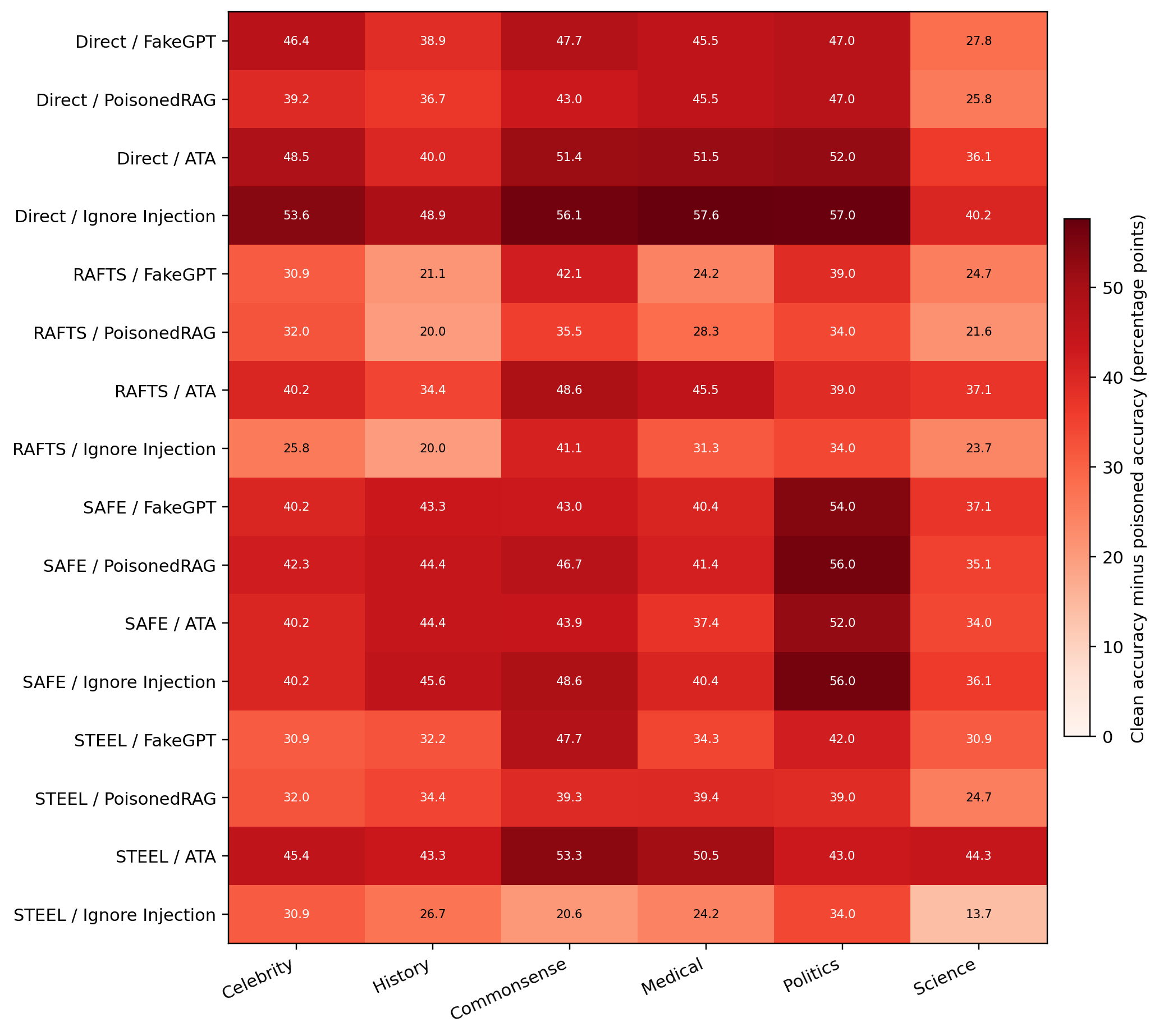}\label{fig:appendix-category-ds}}
  \hfill
  \subfloat[GPT-5.4.]{\includegraphics[width=0.48\textwidth]{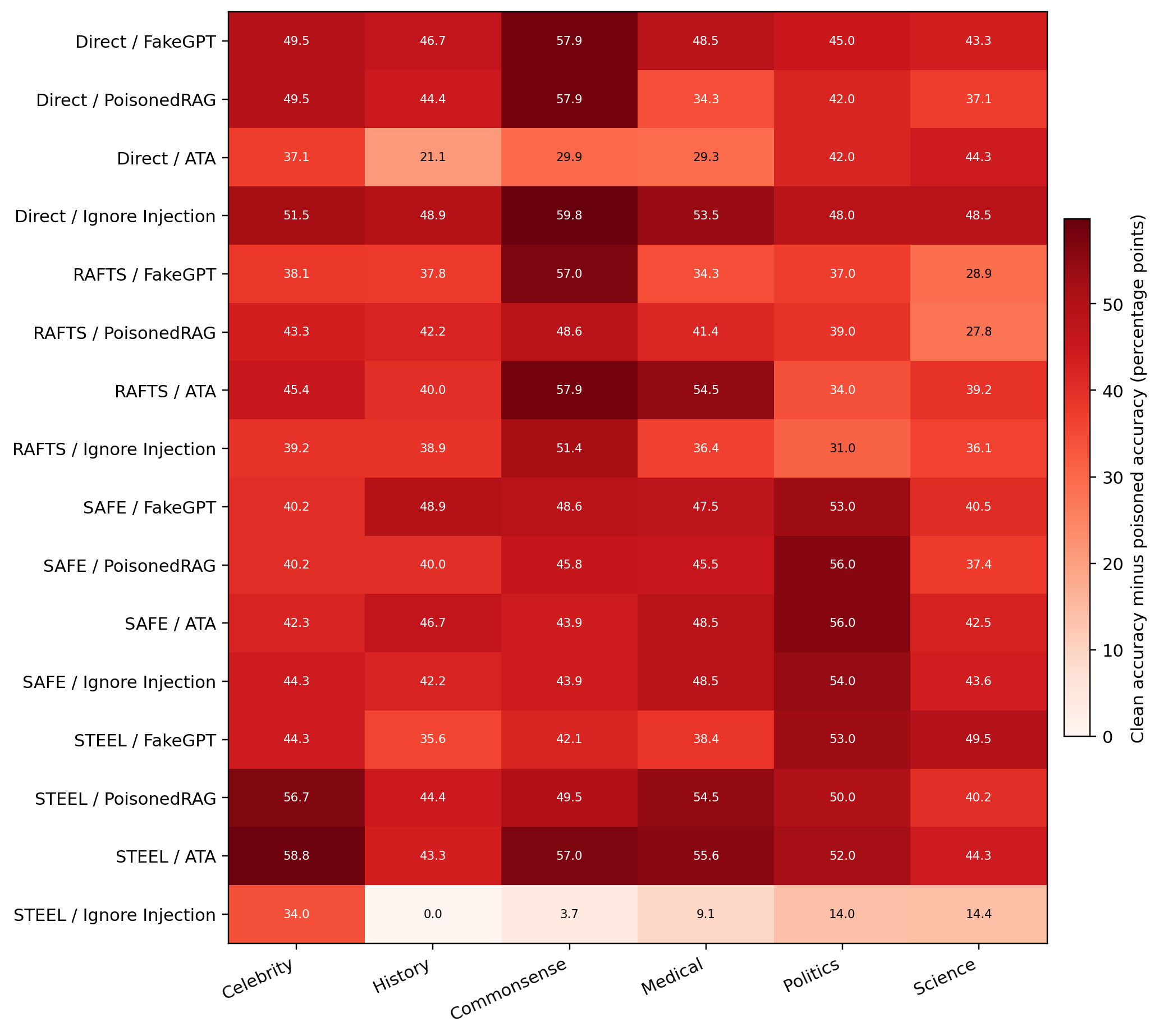}\label{fig:appendix-category-gpt}}
  \caption{Category-level accuracy degradation at 100\% poisoning.}
  \label{fig:appendix-category}
\end{figure*}

Category-level degradation at full poisoning further shows that neither attack effectiveness nor robustness is confined to a single domain. Figure~\ref{fig:appendix-category} measures degradation as clean accuracy minus accuracy at a 100\% poison ratio, so positive values indicate an accuracy loss. ATA is consistently damaging. Under Ignore Injection, STEEL has the smallest loss in every GPT-5.4 category and four DeepSeek-V4-Flash categories; RAFTS is strongest on the remaining two.

GPT-5.4 shows a sharper interaction: Direct and SAFE lose much of their clean accuracy under Ignore Injection, while STEEL degrades comparatively little across all categories. STEEL remains vulnerable to FakeGPT, PoisonedRAG, and ATA, so its advantage is specific to instruction-like evidence. Overall, robustness cannot be inferred from clean accuracy or a single attack.

Table~\ref{tab:efficiency-results} separates robustness from inference cost. On clean evidence, Direct has the lowest TCV (2.00K), followed by STEEL (2.89K), RAFTS (4.28K), and SAFE (8.43K); argument construction and repeated fact verification increase the latter methods' costs without improving clean accuracy. Poisoning raises accuracy-adjusted cost further: Direct's TCV reaches 11.92K, 9.18K, 14.11K, and 69.96K under full FakeGPT, PoisonedRAG, ATA, and Ignore Injection, respectively. Since TCV divides tokens by correct outputs, this growth captures both inference overhead and robustness failure: inexpensive predictions become inefficient when accuracy collapses. GPT-5.4 exhibits the same trade-off.

Beyond exact accuracy and token cost, we examine category-level behavior, subclaim verification, Macro-F1, and ordinal mean score. Each fixed configuration is evaluated once on all 638 claims, with seed controlling evidence selection. Micro-precision, micro-recall, micro-F1, and weighted recall are omitted because they equal exact accuracy in this single-label multiclass setting.

Table~\ref{tab:appendix-clean-category} indicates that clean performance varies by topic and verification procedure. On DeepSeek-V4-Flash, Direct leads on celebrity and medical claims, STEEL on history, commonsense, and science, while Direct, SAFE, and STEEL tie on politics. On GPT-5.4, STEEL leads on celebrity, medical, and science, RAFTS on history and commonsense, and SAFE on politics. The absence of a universal winner shows that aggregate accuracy combines distinct domain-specific strengths.

\begin{table*}[t]
  \centering
  \caption{Clean exact accuracy (\%) by claim category.}
  \label{tab:appendix-clean-category}
  \begin{tabular}{llrrrrrr}
    \toprule
    Model & Method & Celebrity & History & Commonsense & Medical & Politics & Science \\
    \midrule
    \multirow{4}{*}{DeepSeek-V4-Flash} & Direct & 53.3 & 48.0 & 53.9 & 58.9 & 56.5 & 44.8 \\
     & RAFTS & 46.7 & 44.9 & 54.8 & 54.2 & 50.9 & 46.7 \\
     & SAFE & 42.9 & 45.9 & 49.6 & 43.9 & 56.5 & 38.1 \\
     & STEEL & 50.5 & 49.0 & 55.7 & 54.2 & 56.5 & 48.6 \\
    \midrule
    \multirow{4}{*}{GPT-5.4} & Direct & 52.4 & 45.9 & 58.3 & 52.3 & 44.4 & 46.7 \\
     & RAFTS & 47.6 & 54.1 & 60.9 & 56.1 & 44.4 & 46.7 \\
     & SAFE & 44.8 & 48.0 & 50.4 & 49.5 & 56.5 & 45.7 \\
     & STEEL & 55.2 & 52.0 & 55.7 & 57.9 & 50.0 & 52.4 \\
    \bottomrule
  \end{tabular}
\end{table*}

Fine-grained results in Table~\ref{tab:appendix-subclaim} cover the 355 annotated subclaims. Exact accuracy requires the predicted six-way label to match exactly, while ordinal mean score gives partial credit to predictions near the gold label. Let $o(y)\in\{0,1,2,3,4\}$ denote the ordinal index of $y$ in the sequence \emph{false}, \emph{mostly false}, \emph{half true}, \emph{mostly true}, and \emph{true}. For a gold label $y$ and prediction $\hat y$, the score is
\begin{equation}
s(y,\hat y)=
\begin{cases}
1, & y=\hat y,\\
0, & \mathrm{uncertain}\in\{y,\hat y\},\\
\max\!\left(0,1-\frac{|o(y)-o(\hat y)|}{4}\right), & \text{otherwise}.
\end{cases}
\end{equation}
The reported mean score is $100\times\frac{1}{N}\sum_{i=1}^{N}s(y_i,\hat y_i)$. The two metrics therefore separate exact label recovery from approximate veracity estimation.

\begin{table*}[t]
  \centering
  \small
  \caption{Fine-grained subclaim performance (\%). Exact accuracy requires an exact six-way label match; mean score gives partial credit according to ordinal label distance.}
  \label{tab:appendix-subclaim}
  \setlength{\tabcolsep}{1.7pt}
  \begin{tabular*}{\textwidth}{@{\extracolsep{\fill}}lllc *{4}{ccc}@{}}
    \toprule
    Model & Method & Metric & $0\%$ & \multicolumn{3}{c}{FakeGPT} & \multicolumn{3}{c}{PoisonedRAG} & \multicolumn{3}{c}{ATA} & \multicolumn{3}{c}{Ignore Injection} \\
    \cmidrule(lr){5-7}\cmidrule(lr){8-10}\cmidrule(lr){11-13}\cmidrule(lr){14-16}
    & & & & $33\%$ & $67\%$ & $100\%$ & $33\%$ & $67\%$ & $100\%$ & $33\%$ & $67\%$ & $100\%$ & $33\%$ & $67\%$ & $100\%$ \\
    \midrule
    \multirow{8}{*}{DeepSeek-V4-Flash} & Direct & Exact acc. & 73.8 & 60.0 & 53.8 & 24.2 & 63.7 & 53.8 & 23.7 & 53.5 & 42.5 & 27.0 & 67.0 & 56.3 & 8.5 \\
     &  & Mean score & 80.8 & 74.4 & 61.8 & 29.4 & 74.2 & 61.4 & 29.2 & 71.6 & 55.0 & 36.7 & 72.1 & 60.8 & 9.0 \\
    \cmidrule(lr){2-16}
     & RAFTS & Exact acc. & 65.9 & 51.5 & 43.1 & 29.0 & 47.0 & 45.6 & 25.6 & 42.0 & 33.2 & 24.5 & 58.3 & 51.5 & 20.8 \\
     &  & Mean score & 81.9 & 73.4 & 66.7 & 43.2 & 71.5 & 67.0 & 41.5 & 69.7 & 61.2 & 41.9 & 75.8 & 70.4 & 31.2 \\
    \cmidrule(lr){2-16}
     & SAFE & Exact acc. & 71.8 & 60.8 & 51.0 & 26.8 & 59.7 & 49.3 & 25.9 & 55.8 & 45.4 & 30.4 & 68.5 & 59.2 & 10.4 \\
     &  & Mean score & 83.9 & 71.7 & 61.4 & 35.8 & 69.6 & 60.3 & 34.6 & 67.0 & 55.5 & 39.0 & 79.6 & 70.8 & 22.4 \\
    \cmidrule(lr){2-16}
     & STEEL & Exact acc. & 74.4 & 57.5 & 47.6 & 26.8 & 55.2 & 51.3 & 27.0 & 47.0 & 38.3 & 30.4 & 76.3 & 66.2 & 29.3 \\
     &  & Mean score & 85.2 & 74.8 & 64.9 & 38.8 & 73.2 & 66.4 & 39.7 & 72.4 & 60.8 & 40.5 & 87.6 & 81.3 & 60.3 \\
    \midrule
    \multirow{8}{*}{GPT-5.4} & Direct & Exact acc. & 65.1 & 61.4 & 36.1 & 20.3 & 47.9 & 31.3 & 19.7 & 39.2 & 31.3 & 27.0 & 58.0 & 20.3 & 7.0 \\
     &  & Mean score & 72.7 & 71.6 & 55.3 & 31.1 & 64.2 & 47.3 & 28.1 & 64.4 & 50.3 & 36.5 & 66.2 & 23.9 & 8.7 \\
    \cmidrule(lr){2-16}
     & RAFTS & Exact acc. & 51.5 & 19.2 & 18.3 & 14.4 & 14.4 & 14.9 & 14.1 & 10.4 & 9.6 & 11.0 & 18.6 & 18.3 & 20.6 \\
     &  & Mean score & 80.8 & 63.1 & 58.1 & 48.1 & 59.6 & 56.5 & 39.2 & 56.0 & 52.7 & 42.5 & 59.2 & 54.5 & 33.7 \\
    \cmidrule(lr){2-16}
     & SAFE & Exact acc. & 77.7 & 57.2 & 44.2 & 29.6 & 50.4 & 41.4 & 24.8 & 48.7 & 35.2 & 24.2 & 52.7 & 26.8 & 11.5 \\
     &  & Mean score & 87.5 & 66.8 & 53.4 & 36.3 & 59.6 & 49.7 & 32.7 & 58.5 & 43.9 & 31.7 & 63.2 & 33.8 & 16.1 \\
    \cmidrule(lr){2-16}
     & STEEL & Exact acc. & 68.7 & 49.6 & 33.2 & 18.6 & 40.6 & 33.0 & 18.3 & 35.8 & 21.1 & 17.5 & 65.9 & 59.2 & 49.3 \\
     &  & Mean score & 89.0 & 76.6 & 58.9 & 41.5 & 67.5 & 56.3 & 37.0 & 66.5 & 50.6 & 36.4 & 88.6 & 84.4 & 73.8 \\
    \bottomrule
  \end{tabular*}
\end{table*}

Under clean evidence, the best exact subclaim accuracy is 74.4\% for STEEL on DeepSeek-V4-Flash and 77.7\% for SAFE on GPT-5.4. Mean scores are consistently higher than exact accuracy, showing that many errors remain close in ordinal label space. Under full Ignore Injection on GPT-5.4, STEEL retains 49.3\% exact accuracy and a 73.8\% mean score, whereas the other methods achieve at most 20.6\% exact accuracy.

Table~\ref{tab:appendix-overall-metrics} complements exact accuracy with Macro-F1 and ordinal mean score. Macro-F1 weights the per-label F1 values equally, while ordinal mean score distinguishes severe label reversals from adjacent-label errors.

\begin{table*}[t]
  \centering
  \small
  \caption{Overall classification metrics (\%). Macro-F1 weights all six labels equally; mean score gives partial credit according to ordinal label distance.}
  \label{tab:appendix-overall-metrics}
  \setlength{\tabcolsep}{1.7pt}
  \begin{tabular*}{\textwidth}{@{\extracolsep{\fill}}lllc *{4}{ccc}@{}}
    \toprule
    Model & Method & Metric & $0\%$ & \multicolumn{3}{c}{FakeGPT} & \multicolumn{3}{c}{PoisonedRAG} & \multicolumn{3}{c}{ATA} & \multicolumn{3}{c}{Ignore Injection} \\
    \cmidrule(lr){5-7}\cmidrule(lr){8-10}\cmidrule(lr){11-13}\cmidrule(lr){14-16}
    & & & & $33\%$ & $67\%$ & $100\%$ & $33\%$ & $67\%$ & $100\%$ & $33\%$ & $67\%$ & $100\%$ & $33\%$ & $67\%$ & $100\%$ \\
    \midrule
    \multirow{8}{*}{DeepSeek-V4-Flash} & Direct & Macro-F1 & 37.5 & 29.0 & 21.7 & 7.0 & 22.8 & 23.4 & 8.4 & 26.3 & 21.4 & 9.7 & 26.0 & 21.3 & 1.0 \\
     &  & Mean score & 84.4 & 75.2 & 60.1 & 27.7 & 70.6 & 62.9 & 29.4 & 67.4 & 53.1 & 27.0 & 76.3 & 62.9 & 8.8 \\
    \cmidrule(lr){2-16}
     & RAFTS & Macro-F1 & 31.2 & 26.5 & 23.5 & 14.9 & 28.9 & 21.9 & 14.5 & 21.6 & 18.5 & 10.3 & 29.9 & 30.0 & 12.9 \\
     &  & Mean score & 82.4 & 74.2 & 69.4 & 42.0 & 75.6 & 67.2 & 44.6 & 70.6 & 63.3 & 35.5 & 81.0 & 80.2 & 44.8 \\
    \cmidrule(lr){2-16}
     & SAFE & Macro-F1 & 26.9 & 18.0 & 13.7 & 3.1 & 17.7 & 14.5 & 2.4 & 16.9 & 10.9 & 3.6 & 22.7 & 20.1 & 1.0 \\
     &  & Mean score & 81.0 & 68.7 & 61.3 & 34.0 & 68.7 & 61.8 & 34.5 & 65.7 & 55.4 & 37.0 & 77.4 & 74.2 & 22.3 \\
    \cmidrule(lr){2-16}
     & STEEL & Macro-F1 & 34.4 & 27.4 & 23.4 & 10.2 & 25.7 & 21.2 & 10.2 & 22.4 & 16.6 & 11.0 & 33.9 & 31.5 & 20.4 \\
     &  & Mean score & 83.5 & 74.7 & 67.4 & 36.0 & 73.1 & 64.7 & 38.9 & 68.4 & 57.6 & 28.3 & 83.5 & 81.5 & 69.5 \\
    \midrule
    \multirow{8}{*}{GPT-5.4} & Direct & Macro-F1 & 35.7 & 31.8 & 22.6 & 4.1 & 27.9 & 23.6 & 9.0 & 30.3 & 29.2 & 14.1 & 33.4 & 13.6 & 7.8 \\
     &  & Mean score & 77.2 & 73.8 & 43.9 & 19.9 & 64.5 & 56.1 & 28.3 & 69.8 & 59.4 & 37.2 & 66.1 & 20.5 & 13.3 \\
    \cmidrule(lr){2-16}
     & RAFTS & Macro-F1 & 35.8 & 22.1 & 15.9 & 11.4 & 13.0 & 15.6 & 8.9 & 8.9 & 6.3 & 6.7 & 17.2 & 13.9 & 7.4 \\
     &  & Mean score & 82.5 & 70.5 & 65.0 & 53.6 & 64.0 & 64.5 & 48.2 & 61.4 & 58.0 & 51.0 & 65.4 & 61.5 & 37.4 \\
    \cmidrule(lr){2-16}
     & SAFE & Macro-F1 & 28.8 & 19.9 & 14.7 & 2.5 & 13.4 & 11.5 & 3.7 & 10.9 & 7.0 & 3.0 & 16.5 & 7.5 & 1.1 \\
     &  & Mean score & 82.4 & 68.7 & 59.4 & 32.8 & 57.4 & 52.0 & 33.7 & 53.7 & 41.5 & 32.8 & 60.1 & 39.3 & 18.7 \\
    \cmidrule(lr){2-16}
     & STEEL & Macro-F1 & 33.4 & 31.2 & 21.7 & 12.1 & 25.7 & 22.8 & 5.4 & 23.1 & 13.5 & 6.6 & 35.6 & 35.2 & 26.9 \\
     &  & Mean score & 84.7 & 78.2 & 65.1 & 42.6 & 71.3 & 63.3 & 32.2 & 67.1 & 51.1 & 30.1 & 83.7 & 81.7 & 70.7 \\
    \bottomrule
  \end{tabular*}
\end{table*}

Poisoning reduces both Macro-F1 and ordinal mean score, but not always at the same rate. STEEL preserves the strongest Macro-F1 under full Ignore Injection on both backbones (20.4\% and 26.9\%) and also retains the highest ordinal scores (69.5\% and 70.7\%), consistent with its ability to avoid extreme label shifts under this attack.

\section{Conclusion}
We presented GPE, a benchmark and framework for evaluating robust evidence aggregation under controllable GEO-style poisoning. Holding claims and evidence selection fixed while varying attack and contamination distinguishes clean verification ability from robustness. The experiments reveal attack-dependent degradation: ATA is strongest on average, no verifier is consistently robust, and resistance to instruction injection does not imply resistance to content tampering. Residual accuracy reflects model knowledge and credibility heuristics, while low inference cost does not ensure efficiency when accuracy collapses. GPE therefore enables reproducible comparison beyond clean retrieval settings.

\bibliographystyle{IEEEtran}
\bibliography{main}

@article{gpt-4,
  title={Gpt-4 technical report},
  author={Achiam, Josh and Adler, Steven and Agarwal, Sandhini and Ahmad, Lama and Akkaya, Ilge and Aleman, Florencia Leoni and Almeida, Diogo and Altenschmidt, Janko and Altman, Sam and Anadkat, Shyamal and others},
  journal={arXiv preprint arXiv:2303.08774},
  year={2023}
}

@article{deepseek-v3,
  title={Deepseek-v3 technical report},
  author={Liu, Aixin and Feng, Bei and Xue, Bing and Wang, Bingxuan and Wu, Bochao and Lu, Chengda and Zhao, Chenggang and Deng, Chengqi and Zhang, Chenyu and Ruan, Chong and others},
  journal={arXiv preprint arXiv:2412.19437},
  year={2024}
}

@article{toolformer,
  title={Toolformer: Language models can teach themselves to use tools},
  author={Schick, Timo and Dwivedi-Yu, Jane and Dess{\`\i}, Roberto and Raileanu, Roberta and Lomeli, Maria and Hambro, Eric and Zettlemoyer, Luke and Cancedda, Nicola and Scialom, Thomas},
  journal={Advances in Neural Information Processing Systems},
  volume={36},
  pages={68539--68551},
  year={2023}
}

@inproceedings{geo,
  title={Geo: Generative engine optimization},
  author={Aggarwal, Pranjal and Murahari, Vishvak and Rajpurohit, Tanmay and Kalyan, Ashwin and Narasimhan, Karthik and Deshpande, Ameet},
  booktitle={Proceedings of the 30th ACM SIGKDD Conference on Knowledge Discovery and Data Mining},
  pages={5--16},
  year={2024}
}

@article{poisonedrag,
  title={Poisonedrag: Knowledge corruption attacks to retrieval-augmented generation of large language models},
  author={Zou, Wei and Geng, Runpeng and Wang, Binghui and Jia, Jinyuan},
  journal={arXiv preprint arXiv:2402.07867},
  year={2024}
}

@article{fakegpt,
  title={FakeGPT: fake news generation, explanation and detection of large language models},
  author={Huang, Yue and Sun, Lichao},
  journal={arXiv preprint arXiv:2310.05046},
  year={2023}
}

@inproceedings{scr,
  title     = {Combating Knowledge Corruption in Agent Systems: A Byzantine-Tolerant Secure Collaborative RAG Framework},
  author    = {Wang, Zhaoqi and He, Daqing and Zhang, Zijian and Liu, Ye and Liu, Jiamou and Zeng, Zhirui and Qin, Zhan and Li, Zhen and Li, Xin and Yao, Hongwei and An, Jincheng and Liu, Yong and Li, Yi and Sun, Qi and Liu, Xiulei and Zhu, Liehuang},
  booktitle = {Proceedings of the ACM Web Conference 2026},
  series    = {WWW '26},
  year      = {2026},
  publisher = {ACM},
}

@article{fakenews_survey,
  title={A survey of fake news: Fundamental theories, detection methods, and opportunities},
  author={Zhou, Xinyi and Zafarani, Reza},
  journal={ACM Computing Surveys (CSUR)},
  volume={53},
  number={5},
  pages={1--40},
  year={2020},
  publisher={ACM New York, NY, USA}
}

@inproceedings{f3,
  title={Fighting fire with fire: The dual role of LLMs in crafting and detecting elusive disinformation},
  author={Lucas, Jason and Uchendu, Adaku and Yamashita, Michiharu and Lee, Jooyoung and Rohatgi, Shaurya and Lee, Dongwon},
  booktitle={Proceedings of the 2023 Conference on Empirical Methods in Natural Language Processing},
  pages={14279--14305},
  year={2023}
}

@inproceedings{teller,
  title={Teller: A trustworthy framework for explainable, generalizable and controllable fake news detection},
  author={Liu, Hui and Wang, Wenya and Li, Haoru and Li, Haoliang},
  booktitle={Findings of the Association for Computational Linguistics: ACL 2024},
  pages={15556--15583},
  year={2024}
}

@inproceedings{rafts,
  title={Retrieval augmented fact verification by synthesizing contrastive arguments},
  author={Yue, Zhenrui and Zeng, Huimin and Shang, Lanyu and Liu, Yifan and Zhang, Yang and Wang, Dong},
  booktitle={Proceedings of the 62nd Annual Meeting of the Association for Computational Linguistics (Volume 1: Long Papers)},
  pages={10331--10343},
  year={2024}
}

@article{safes,
  title={Long-form factuality in large language models},
  author={Wei, Jerry and Yang, Chengrun and Song, Xinying and Lu, Yifeng and Hu, Nathan and Huang, Jie and Tran, Dustin and Peng, Daiyi and Liu, Ruibo and Huang, Da and others},
  journal={Advances in Neural Information Processing Systems},
  volume={37},
  pages={80756--80827},
  year={2024}
}

@article{steel,
  title={Re-search for the truth: Multi-round retrieval-augmented large language models are strong fake news detectors},
  author={Li, Guanghua and Lu, Wensheng and Zhang, Wei and Lian, Defu and Lu, Kezhong and Mao, Rui and Shu, Kai and Liao, Hao},
  journal={arXiv preprint arXiv:2403.09747},
  year={2024}
}

@article{adsent,
  title={Robust Fake News Detection using Large Language Models under Adversarial Sentiment Attacks},
  author={Tahmasebi, Sahar and M{\"u}ller-Budack, Eric and Ewerth, Ralph},
  journal={arXiv preprint arXiv:2601.15277},
  year={2026}
}

@inproceedings{liar,
  title={“liar, liar pants on fire”: A new benchmark dataset for fake news detection},
  author={Wang, William Yang},
  booktitle={Proceedings of the 55th annual meeting of the association for computational linguistics (volume 2: short papers)},
  pages={422--426},
  year={2017}
}

@dataset{politifact,
  author = {Misra, Rishabh},
  year = {2022},
  month = {09},
  pages = {},
  title = {Politifact Fact Check Dataset},
  doi = {10.13140/RG.2.2.29923.22566}
}

@inproceedings{fever,
    title = "{FEVER}: a Large-scale Dataset for Fact Extraction and {VER}ification",
    author = "Thorne, James  and
      Vlachos, Andreas  and
      Christodoulopoulos, Christos  and
      Mittal, Arpit",
    editor = "Walker, Marilyn  and
      Ji, Heng  and
      Stent, Amanda",
    booktitle = "Proceedings of the 2018 Conference of the North {A}merican Chapter of the Association for Computational Linguistics: Human Language Technologies, Volume 1 (Long Papers)",
    month = jun,
    year = "2018",
    address = "New Orleans, Louisiana",
    publisher = "Association for Computational Linguistics",
    url = "https://aclanthology.org/N18-1074/",
    doi = "10.18653/v1/N18-1074",
    pages = "809--819"
}

@article{gossipcop,
  title={Fakenewsnet: A data repository with news content, social context, and spatiotemporal information for studying fake news on social media},
  author={Shu, Kai and Mahudeswaran, Deepak and Wang, Suhang and Lee, Dongwon and Liu, Huan},
  journal={Big data},
  volume={8},
  number={3},
  pages={171--188},
  year={2020},
  publisher={SAGE Publications Sage CA: Los Angeles, CA}
}

@inproceedings{check_covid,
  title={Check-COVID: Fact-checking COVID-19 news claims with scientific evidence},
  author={Wang, Gengyu and Harwood, Kate and Chillrud, Lawrence and Ananthram, Amith and Subbiah, Melanie and McKeown, Kathleen},
  booktitle={Findings of the Association for Computational Linguistics: ACL 2023},
  pages={14114--14127},
  year={2023}
}

@misc{deepseekv4,
      title={DeepSeek-V4: Towards Highly Efficient Million-Token Context Intelligence},
      author={DeepSeek-AI},
      year={2026},
}

@misc{gpt54,
  author       = {OpenAI},
  title        = {Introducing {GPT-5.4}},
  year         = {2026},
  month        = {March},
  howpublished = {\url{https://openai.com/index/introducing-gpt-5-4/}},
  note         = {Accessed: 2026-03-05}
}

\end{document}